\documentclass[10pt,manuscript]{aastex}

\slugcomment{To Appear in ApJL}
\received{July 31, 2009}
\revised{January 08, 2010}
\accepted{January 14, 2010}

\shorttitle{Proper-Motion Study of the Circumstellar Shell of HD 179821} 
\shortauthors{Ferguson \& Ueta}

\begin{document}
 
\title{Differential Proper-Motion Study of the Circumstellar Dust Shell\\
of the Enigmatic Object, HD 179821} 

\author{Brian A.\ Ferguson\altaffilmark{1}, Toshiya Ueta}

\affil{Department of Physics and Astronomy,
University of Denver, 
Denver, CO 80208}

\altaffiltext{1}{Current address: Space Telescope Science Institute,
3700 San Martin Dr., Baltimore, MD 21218} 
\email{bferg@stsci.edu}

\begin{abstract}
HD179821 is an enigmatic evolved star that possesses
 characteristics of both a post-asymptotic giant branch star and a
 yellow hyper-giant, and there has been no evidence that unambiguously
 defines its nature. These two hypotheses are products of an
 indeterminate distance, presumed to be 1 kpc or 6 kpc.  
 We have obtained the two-epoch Hubble Space Telescope WFPC2 data of its
 circumstellar shell, which shows multiple concentric arcs extending out
 to about $8\arcsec$.
 We have performed differential proper-motion measurements on distinct
 structures within the circumstellar shell of this mysterious star in hopes
 of determining the distance
 to the object, and thereby distinguishing the nature of this enigmatic stellar source.
 Upon investigation, rather than azimuthal radially symmetric expansion,
 we discovered a bulk motion of the circumstellar
 shell of ($2.41 \pm 0.43$, $2.97 \pm 0.32$) mas yr$^{-1}$.  
 This corresponded to a translational ISM flow of (1.28$\pm$0.95,
 7.27$\pm$0.75) mas yr$^{-1}$ local to the star.
 This finding implies that the distance to HD 179821 should be rather
 small in order for its circumstellar shell to preserve its highly intact
 spherical structure in the presence of the distorting ISM flow, therefore favoring the proposition that HD 179821 is a
 post-AGB object. 
\end{abstract}


\section{Introduction}

The understanding of the nature of the evolved star HD 179821 (IRAS
19114$+$0002; AFGL 2343; hereafter HD 179821) is presently ambiguous. 
At present, there exist two plausible distance estimates for the object,
hence two plausible, yet significantly different, masses and
luminosities.  

Metal abundances imply that HD 179821 has gone through the third dredge-up,
and if so its distance would be roughly 1 kpc based on an assumed luminosity \citep{rh99}.  
At this distance, this star is most likely a post-asymptotic giant
branch (post-AGB) star with a luminosity near $1.6 \times 10^{4}$
L$_{\odot}$, implying an initial mass of 1 M$_{\odot}$ \citep{hkv89}. 
The other competing theory, inferred from stellar expansion velocities, 
galactic disk models \citep{rh99}, maser line observations, and an
infrared (IR) excess \citep{jmbdh94}, is that HD 179821 lies at 6 kpc.  
At this distance, this star is most likely a yellow hyper-giant with a
high luminosity near $3.1 \times 10^{5}$ L$_{\odot}$ \citep{hawkins95}
and an implied mass near 30 M$_{\odot}$ \citep{schaller92}.
Even the spectral type of the star is still open to scrutiny.
HD 179821 was observed to be a G5Ia type star \citep{hkv89}.
However, the spectroscopic surface temperature of 6800 K (which implies
an F5 star) derived by \citet{zacs96} significantly differs from what is 
expected for the G5Ia star (5100 K). 

Despite the discrepancies, it is indisputably agreed that HD 179821 is a
high luminosity object with a detached cool dust shell, which is
suggestive of mass loss. 
This dust shell is formed during the late stages of evolution, typical
for stars characterized by strong stellar winds (e.g.\
\citealt{waters96}). 
Large dusty envelopes have rarely been observed around
yellow hyper-giants, but when they do, they show themselves by an
infrared excess \citep{jmbdh94}. 
In HD 179821's particular case, its circumstellar dust shell has been
imaged in the thermal dust emission at mid-IR \citep{jura99,ueta01}.

Five separate CO observations
(\citealt{zd86,likkel87,vtw93,fong06,cas07}), and three photospheric
line analyses \citep{zacs96,rh99,kipper08} suggest that HD 179821  
possesses a heliocentric radial velocity of $v_{\rm r} = 84.8 \pm 1.4$ km s$^{-1}$,
as well as a large outflow velocity, $v_{\rm exp} = 35 \pm 2$ km s$^{-1}$.  
The shell expansion velocity is about double the expected wind velocity
of typical post-AGB type stars.  
Therefore the larger of the two distances is suggested, (inferring that this
large shell expansion velocity suggests a more massive super-giant).

Following the discovery of the circumstellar shell of HD 179821 in the
dust-scattered star light in the optical \citep{umb00} and in the
near-IR, \citet{gledhill01}  proceeded to map the shell in
the OH 1667- and 1612-MHz maser lines, showing that the emission arose
from a relatively think shell whose OH dissociation characteristic is
consistent with OH shell models at 6 kpc.
In addition, the observed oxygen to carbon ratio of 2.6 O/C for HD
179821 is significantly large to infer that it is a low- to
intermediate-mass object \citep{rh99}.
However, this star's metal composition shows an overabundance of
s-process elements \citep{zacs96,rh99,kipper08}, which suggests the
post-AGB nature of the star.

To make the matter more complicated, \citet{rh99} alternatively
suggested that HD 179821 is located in the Galactic disk about 4 kpc
away based on the presence of spectral lines at 5780.41 and 5797.03 \AA,
(originating from the intervening diffuse interstellar clouds in the
vicinity of the star), and the comparison between the velocity of the clouds
and the radial velocity structure of the Galactic disk.
Distance determination is generally one of the most difficult astronomical 
tasks, and this quantity remains elusive for HD 179821.

\section{Differential Proper-Motion Measurements}

Differential proper-motion measurements of the circumstellar shell
provide a direct means for distance determination, under the assumption
that 
(1) the rate of expansion is constant over the interval of the two
measurements,
(2) the expansion is understood as a radially symmetric motion,
and
(3) the line-of-sight expansion velocity measured by some other means is
equal in magnitude to the translational expansion derived from the
differential proper-motion measurements

In the past, this methodology was tested and proven successful in
studying the circumstellar shells of, for example: $\eta$ Car
\citep{currie96,morse01} and the Cygnus Egg Nebula \citep{umm06}.   
Because HD 179821 was observed by the {\sl Hubble Space Telescope}
({\sl HST}) on 1997 July 12 (GO-6737) in the previous imaging study
\citep{umb00}, its shell should have expanded by 
anywhere from 12 to 72 milli-arcseconds over the course of roughly $10$
yr at the known expansion rate of 35 $\pm$ 2 km s$^{-1}$ if the object is
located at a distance ranging from 6 to 1 kpc, respectively.
Because these expected amounts of translational expansion are detectable
with the {\sl HST}, we carried out the second epoch imaging of the
object to perform the differential proper-motion
measurements on 2007 July 6 (GO-10837), in order to provide another
independent estimate of the distance to the object.

\section{Observations and Data Reduction}

For the present work we used two sets of images with the continuum F547M
medium bandpass filter of the WFPC2 camera aboard the {\sl HST}.  
HD 179821's shell is small enough ($\sim8\arcsec$ diameter) to
be completely observed within the Planetary Camera (PC) chip
frame of WFPC2, which affords the highest resolution of the WFPC2 instrument. 
The first epoch observations used for the initial condition of the
differential proper-motion study were taken on 1997 July 12.  
The second epoch data were taken on 2007 July 6. 
Both sets, taken during a single orbit, include a long 30 sec exposure
time. 

 The interval between the observations of HD 179821 provides
for 10 years of stellar motion (corresponding to half a pixel at a 6 
kpc distance and a constant velocity of 35 $\pm$ 2 km s$^{-1}$, coupled
with the PC pixel scale of $0\farcs0455$ pixel$^{-1}$).
To attain a clear enough image for our proper-motion measurements, we
used multiple dithered exposures.
Multiple exposures are needed to account for the difficulty in
distinguishing between real sources and common image defects
(e.g.\ cosmic rays).  
Also because these dithered images overlap, if the images
are shifted a sub-pixel amount and recombined, then a sub-pixel
structure can be restored (which is very important for sub-pixel level 
differential proper-motion analyses).  
The limit of this is a quarter of a pixel once the dithered images have
been brought to a mosaic, which is possible with the difference-squared
correlation method employed by \citet{currie96}, \citet{morse01} and
\citet{umm06}. 

For data reduction, we used the standard set of WFPC2 calibration
protocols as provided and defined in IRAF\footnote{IRAF is distributed
by NOAO, which is operated by the AURA, under cooperative agreement with
the NSF, as a standard image reduction tool.} STSDAS\footnote{Space
Telescope Science Data Analysis System, Version 3.9 copyright 2003
Association of Universities for Research in Astronomy, Inc. (AURA).}.  
Most image defects contained in the dithered images
were corrected using sky mask subtraction and cosmic ray rejection tasks
included in the dithered image combining task, MULTIDRIZZLE (Version 2.3:
\citealt{koe02}), in STSDAS.
However, after this cleaning process there remained an anomaly 
that had to be manually corrected prior to drizzling.  
There was a region of slightly elevated pixel values in the middle of
the PC chip, spanning roughly 140 pixels wide and extending for the
entire 800 pixel length along the readout direction of the PC chip
(i.e.\ across the circumstellar shell). 
This is caused by the bright central star being overexposed for 30 sec.
As the defect was relatively constant in the affected region, we were
able to estimate the amount of the pixel value elevation away from the
central star and the circumstellar shell. 
After correcting for this anomaly, the ambient background is leveled to
approximately a zero pixel value. 
This was performed on each of the separate images taken with the long
exposure time for the epochs. 

Although default distortion and rotation correction files called by
MULTIDRIZZLE are accurate for a general instrumental distortion,
these plate solutions called in the header of our data files were not
quite good enough for our scientific goals.  
To measure proper motion from the two-epoch data, the images must be 
aligned to a sub-pixel level precision over these two epochs. 
In addition to the drizzling task, we performed an additional distortion
correction and alignment using background stars. 
A total of 22 point-sources that appear in the PC chip of both sets of
long exposure images were identified as references.
The point-spread-function (PSF) of these reference background stars were
fit by Gaussian profiles to obtain accurate positions, and then
instrumental distortions were corrected with the GEOMAP and GEOTRAN
tasks of STSDAS.  
Due to the fact that each long exposure image would have a similar
distortion error as their corresponding short exposure image, we used
the plate solution of the short exposure, applied it to the
long, and then ran a supplemental final distortion correction on the
long exposure data set.  
The distortion correction was carried out with
an accuracy of 0.08 pixels (3.6 mas).

\section{Results}

After aligning two images, we carried out differential proper-motion
measurements following the method elucidated and implemented previously
\citep{currie96,morse01,umm06}.   
The nebula structures are seen via the star light scattered by dust
grains in the circumstellar shell. 
In order to enhance the shell structure, we subtracted
an azimuthally-averaged radial profile centered at the position of the
star.
While the general structural characteristic of the nebula is its
multiple concentric shells, there are a number of distinct local
structures with which we can measure the amount of differential proper 
motion in the shell over the 10 year interval.  
In total, 33 distinct positions were identified along the outer rings.   

\begin{figure}
 \begin{center}
 \includegraphics[width=2.2in]{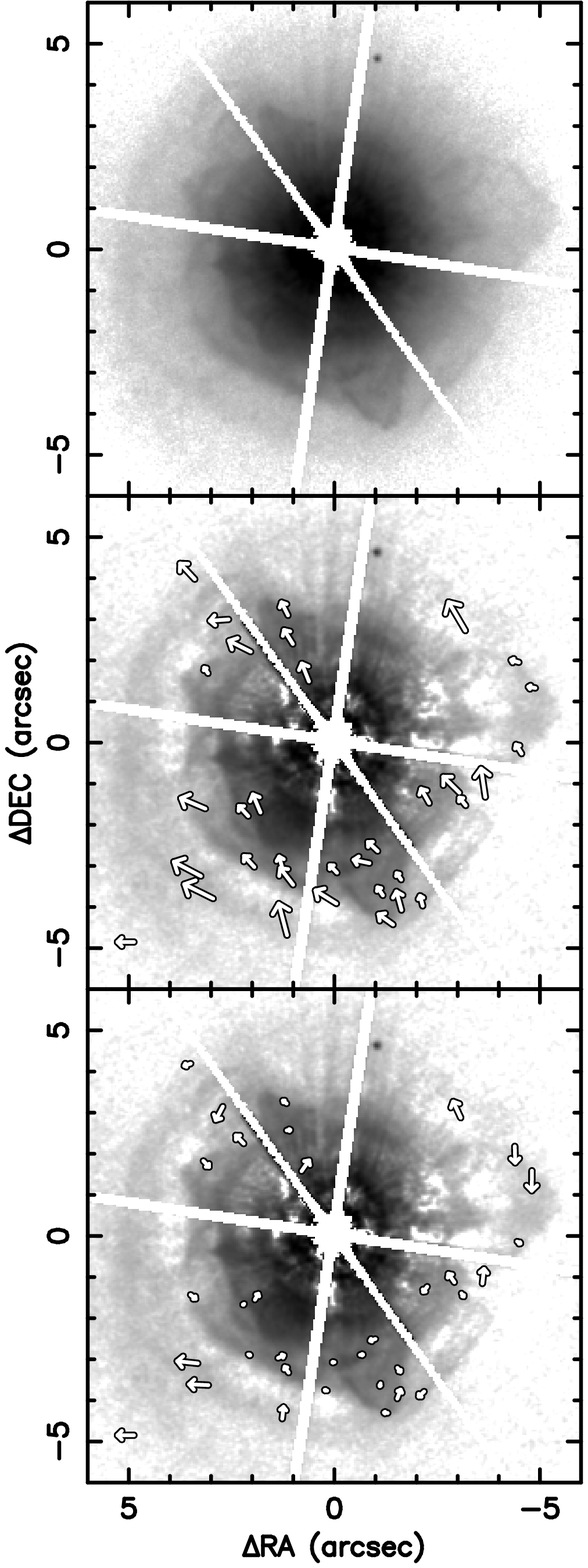}
 \end{center}
 \caption{\label{map}%
 [Top] Cleaned, anomolly-corrected, masked image of HD 179821 centered at
 the position of the central star at the epoch 1. North is up, east to
 the left. Tickmarks indicate RA and DEC offsets in arcseconds.  
 [Middle] azimuthal-average-subtracted image of HD 179821 overlaid with
 arrows indicating the measured differential proper-motion of the 33
 local structures suggestive of the shell's bulk motion.
 [Bottom] same as middle but overlaid with arrows indicating the residual
 differential proper-motion of the 33 local structures after the suspected
 bulk motion is removed.  The arrow on the bottom left shows a vector
 corresponding to a measured one-pixel shift.}    
\end{figure}

At this stage we were forced to remove any significantly saturated pixels in
the image as saturated pixels throw off the matching process involved in 
differential proper-motion measurements. 
In the final drizzled images there existed two narrow diffraction spikes
introduced by {\sl HST} instrumental optics extending the length of the
image.  
There was also a short, centered, diamond-shaped saturation spike
arising from the over exposed central star.
A mask was created and both effects were subsequently removed (see the
top frame of Figure \ref{map}).

To determine the amount of shift, we first defined square image sections
centered at each of these well-defined local structures in the epoch 1
image. 
The size of these image sections has to be large enough (typically
$\sim10$ pixels) so that the segmented structures can be uniquely
identified.   
In the second epoch image we defined a search box that is three times
larger than the image section centered at the epoch 1 position of the
local structures. 
In the epoch 2 search box, we cut out an image section that is the
same size as the epoch 1 image section and compared the two by means of
cross-correlation. 
We did this by computing the inverse of the sum of the squares of the
difference between the corresponding pixel values in the two image
sections. 
Since this value represents the ``correlation'' between the two image
sections, it tends to be large if the two image sections capture the same
local structure (hence, the two are similar), but small if the two are
different. 
We repeated this operation while the epoch 2 image section was moved
around inside the search box, and at the end we searched for the optimal
shift that is represented by the maximum correlation. 
This shift is the measured shift of the local structure between the
first and second epochs.
This procedure was repeated for all of the 33 local structures that we
selected.
The middle frame of Figure \ref{map} shows these measured shifts by
arrow symbols, whose lengths correspond to the magnitudes of the shift
and angles to the directions.

Surprisingly, the measured shifts of the local structures in the
circumstellar shells of HD 179821 indicate that the whole nebula
has exhibited a bulk motion toward the northeast direction in the plane
of the sky.
Under the assumption that the shell expands radially as its mass is
ejected from the central star, we expect to see centrosymmetrical
relative motion of sub-structures in the nebula.  
Still assuming that expansion should have been radially symmetric, we
can search for a respective shift of the second epoch image (which is
due to the bulk motion of the shell) that will recover residual shifts
that are more radially symmetric---when the measured shifts of the
local structures account for this induced shift of the second epoch image. 

To quantitatively describe the bulk motion and to find the true differential shell
expansion, we iteratively shifted the epoch 2 image and performed the
differential proper-motion measurement procedure until the vectors
representing residual motion of the sub-structures were pointed in the
most radially outward directions. 
After this exercise, we were able to determine that the bulk motion of
the circumstellar shell was (2.41$\pm$0.43, 2.97$\pm$0.32) mas yr$^{-1}$. 
The residual shifts of the local structures are shown in the bottom
frame of Figure \ref{map}.
Although this search did converge, the resultant residual shift vectors
do not exhibit an obvious symmetric shell expansion.

\section{Discussion}

\subsection{Distance}

The distance estimate based on the differential proper-motion
measurement of the circumstellar shell relies on our ability in
(1) determining the magnitudes and directions of shell expansion  
from the two-epoch images and (2) relating them to the known expansion 
velocity of the shell under the assumption of an isotropic 
shell expansion. 
Our analysis for HD 179821, however, did not conclude a symmetric shell
expansion: it instead yielded residual shift vectors that show
multitudes of lengths and directions. 
If we take the average of these resultant expansion vectors of
individual shell structures to relate to 35 km s$^{-1}$, the distance
to HD 179821 is estimated to be $3.91 \pm 3.23$ kpc. 
The large error reflects a large scatter in the length of the resultant 
expansion vectors.
Because of the large error, this result cannot address the dichotomy of 
the distance estimates for HD 179821.

The 33 distinct shell structures that were used to follow the shell
expansion are preferentially located near the periphery of the reflection
nebulosity.  
This is because no individual structures were identified in the
central part of the shell to follow the shell motion because of the
bright PSF.
The most recent molecular observations of the circumstellar shells of HD 
179821 have resolved a concentric structure of $\sim 2\arcsec$ radius,
which is responsible for the bulk of CO and HCN emission
\citep{cas07,qbc08}.     
Ideally, the 35 km s$^{-1}$ expansion velocity has to be related to 
the expansion of this $2\arcsec$ radius molecular shell in the
differential proper-motion analysis.
Due to the bright central source, we were only able to follow the shell motion in regions $3\arcsec$
to $5\arcsec$ away from the central point of the object. 

It is, therefore, possible that the shell expansion near the periphery
of the shell, represented by the residual expansion vectors recovered in
our analysis, has been taking place at velocities lower than 35 km s$^{-1}$, having 
suffered from pile-up and slow-down of the wind material in front of
the interface regions between the stellar wind and interstellar medium 
(ISM) as predicted numerically by, e.g.,\ \citet{steff98}.  

There is another possibility for the apparently slowed outer shells, although highly doubtful.
Instead of being slowed by ISM interaction, it is possible that the outer shells were actually
ejected during an earlier AGB phase at a lower velocity than the shells of mass most recently ejected.  
This scenario would imply two different spherically observed shell expansion velocities.  This smaller outflow velocity
would presumably be transparent in a molecular line study of radial expansion, showing two Doppler-shifted peaks about center: one for the smaller
on-coming velocity of the outer shell and one for the receding velocity---neither of which are observed.  Even the most recent high-resolution mapping in molecular lines bear no
indication that HD 179821's shell is expanding at anything other than a spherically consistent singular velocity.  In fact
the only small deviation that exists arises from
the internal structure of the shell, instead of the would-be expected outer edges \citep{fong06,cas07,qbc08}.  

With the shell expansion actually taking place at lower velocities due to the probable ISM interaction, lies the implication 
that the free-expansion stellar wind shell is actually of
$\sim 2\arcsec$ to $3\arcsec$ radius in size and is surrounded by the
region of compressed wind material in which the wind velocity is
lowered.   
This interpretation naturally explains the fact that the resultant  
expansion vectors are small and appear to be oriented in seemingly
random directions. 
Such non-radial internal motion of the shell has also been reported 
in molecular observations \citep{cas07,qbc08}.  
Thus, being that it is likely that the shell motion has been taking place at 
velocities lower than 35 km s$^{-1}$, the distance estimate obtained above (using this shell expansion value) 
would mean the upper limit.  Regardless, using direct proper-motion analysis the distance to HD 179821 can still be 1 or 6 kpc, and
therefore, is inconclusive.

\subsection{The Bulk Motion of the Shell}

The detected bulk motion of the shell amounts to 0.85 pixel, about twice
larger than the amount of the residual expansion vectors. 
There exist three obvious possibilities as the source of this bulk
motion. 
First is proper motion of the central star, which should induce the
bulk motion if the shell is co-moving with the central star. 
Second is motion of the ISM relative to HD 179821, which would push
the shell in the direction of the ISM flow in the plane of the sky.
Third is synchronized proper motion of all 22 nearby stars used to
align the two epoch images. 
The third possibility requires all 22 stars having been moving in the
same direction by the same amount over the past 10 years.
We do not consider this feasible, as these stars demonstrate absolutely no 
differential motion between each other, after the two epoch frames have been aligned.
Therefore these stars, if moving, must have a perfectly synchronized velocity in respect to their distance so as to seem all moving the exact same amount from the perspective of the Earth-based astronomer.  Furthermore, no such homogeneous motion of nearby stars has been reported in, e.g., the 
Naval Observatory Merged Astrometric Dataset \citep{nomad}. 

The central star's proper motion is known to be (1.13$\pm$0.85,
-4.30$\pm$0.68 ) mas yr$^{-1}$ \citep{vanl07}.
Hence, the shell is apparently moving into the almost opposite direction 
in the plane of the sky.
Therefore, in the frame of the central star, there is a translational
ISM flow local to HD 179821 at (1.28$\pm$0.95, 7.27$\pm$0.75) mas
yr$^{-1}$, which is nearly due north in the plane of the sky.  
Since we have two-epoch imaging data, we should be able to confirm the
{\sl Hipparcos\/}-measured proper-motion of the star using our data. 
It turned out to be challenging because the central star was saturated
even in the shortest exposure (4 sec) frames during the second-epoch 
observations.  
Nevertheless, we attempted to measure proper motion of the star through
pinpointing the location of the star by tracing the PSF spikes, (i.e.\
determining where the PSF spikes converge), because they
should emanate from the central star.  

For the epoch 1 data, in which the central star was not saturated in the
shortest exposure (0.01 sec) images, we can determine if the PSF spikes
would really pinpoint the position of the star. 
The peak centroid position of the observed star and the location where 
the PSF spikes cross were offset by 0.73 pix, which is larger than both
of the centroiding error and the error in the PSF-spike analysis.
Hence, this PSF-spike analysis would not determine the location of the
central star at the astrometric accuracy that we would prefer for our purposes.  
However, if there is no time and on-chip-position dependent
effects in the way the PSF 
spikes are caused by the central star the relative shift of the position
of the central star determined from the PSF-spike analysis in the two
epochs would represent proper motion of the star.

The PSF-spike analysis done for both epochs yielded proper motion of
the star to be ($-0.08\pm 1.58$, $-0.53\pm 6.27$) mas yr$^{-1}$.
The large errors arose from the accuracy of linear-fitting for the
PSF spikes. 
This means that within the error of the analysis the central star did
not exhibit any appreciable amount of proper motion over the 
two epochs; however this error includes the {\sl Hipparcos\/}-measured 
proper-motion of the star.  (It should however be noted that the  
{\sl Hipparcos\/}-measured 
proper-motion is only valid if the star is in range.)

Since the validity of this PSF-spike method to locate the position of
the star has never been documented even by the WFPC2 instrument team, we 
have no way to access how viable our analysis is.
In any event, based on the available information, we conclude that 
our two-epoch data of HD 179821 imply the existence of a translational
ISM flow local to the object at least ($2.41 \pm 0.43$, $2.97 \pm 0.32$)
mas yr$^{-1}$ and at most ($1.28 \pm 0.95$, $7.27 \pm 0.75$) mas
yr$^{-1}$, dependent on the true amount of proper motion of the central
star. 
These values translate to $18 \pm 2$ km s$^{-1}$ to $35 \pm 4$ km
s$^{-1}$ at 1 kpc and  $109 \pm 10$ km s$^{-1}$ to $210 \pm 22$ km
s$^{-1}$ at 6 kpc.

\subsection{ISM Flow around HD 179821}

The observed bulk motion of the shell around HD 179821 indicates
the presence of an ISM flow local to the object irrespective of its
suggested distances. 
Recently, similar ISM flows have been detected around other mass-losing 
stars based on the discrepancy between the direction of
the central star's space motion vector and the orientation of the bow
shock structure at the interface between the stellar winds and the ISM
inferred from far-IR images \citep{ueta08,ueta09,u09}. 
HD 179821's case does not resemble other ISM flow cases in that the
shell structure of the object does not show any obvious signatures of
shock interactions expected at the wind-ISM interface regions in the
optical through mid-IR \citep{jura99,umb00,gledhill01,ueta01,bum06}. 
Other than this obvious problem, the present case is similar to the case
of an asymptotic giant branch star, R Cas, for which the wind-ISM
boundary appear rather circular as opposed to parabolic that is typical
of a bow shock due to the inclination of the wind-ISM interface regions
($22^{\circ}$ with respect to the line of sight, \citealt{u09}).  

The ISM flow detected in the present analysis is only the translational
component of the space motion vector of the ISM flow.
The radial velocity of the star, based on the photospheric line analysis,
is known to vary but is converged to the heliocentric value of 85 km
s$^{-1}$ \citep{zacs96,rh99,kipper08}.
Molecular observations of the shell  yielded similar values
\citep{zd86,likkel87,vtw93,fong06,cas07}.
Therefore, there is no evidence indicating that the shell and star
are moving differently along the radial direction.
Hence, there is no ISM flow local to HD 179821 in the radial direction.  

Prior to this study, no evidence existed of such an ISM flow in the
vicinity of HD 179821.
Even the most recent high-resolution mapping in molecular lines have
indicated that HD 179821's shell is expanding fairly spherically
symmetrically except for some deviation from circular symmetry seen in
the internal structure of the shell \citep{fong06,cas07,qbc08}.  
Hence, if this ISM flow exists the circumstellar shell must be able to
withstand the flow to maintain its more-or-less spherically symmetric
characteristic.
This is more so, given that the interaction between the shell of HD 179821 
and the local ISM flow is viewed edge-on.
Therefore, the fact that HD 179821's shell does not show any apparent
bow shock, and maintains spherical symmetry, is consistent with the expectation that the
interaction between the shell and the ISM, if exists, must be benign.  

The internal pressure inside the shell can be estimated as follows.
Since the shell has a large physical thickness in front of the bulk of
the molecular-line emitting shell (of $\sim 2\arcsec$ radius) to the
edge of the reflection nebulosity (of $\sim 5\arcsec$ radius), we assume
that the shell is in the energy-conserving expansion phase.
Then, the internal pressure of the shell, $p_{\rm sh}$, can be estimated
by relating the internal energy of gas in the shell and the kinetic
energy of the wind:
\begin{eqnarray}
 p_{\rm sh} = \frac{\frac{1}{2} M_{\rm sh} v_{\rm w}^2}{2 \pi R_{\rm
  sh}^3}
\nonumber
\end{eqnarray}
where $M_{\rm sh}$ is the shell mass, $v_{\rm w}$ is the wind
velocity, and $R_{\rm sh}$ is the radius of the shell (e.g.\
\citealt{lc99}). 
Since $M_{\rm sh} \ge \dot{M} t_{\rm sh}$, where $\dot{M}$
is the rate of mass loss ($\sim 3 \times 10^{-3}$ M$_{\odot}$ yr$^{-1}$; 
\citealt{cas07}) 
and $t_{\rm sh}$ is the dynamical age of the shell (i.e.\ the wind
crossing time, $R_{\rm sh}/v_{\rm w}$), 
\begin{eqnarray}
 p_{\rm sh} \ge \frac{\dot{M} v_{\rm w}}{4 \pi R_{\rm sh}^2}
\nonumber
\end{eqnarray}
For $v_{\rm w} = 35$ km s$^{-1}$ and $R_{\rm sh} = 5\arcsec$, $p_{\rm sh}
\ge 9.4 \times 10^{-7}$ and $2.6 \times 10^{-8}$ dyn cm$^{-2}$ for the shell at
1 and 6 kpc away, respectively.

On the other hand, the ISM pressure impinging on the wind-ISM interface 
regions, $p_{\rm ISM}$, can be approximated by
\begin{eqnarray}
 p_{\rm ISM} = \frac{1}{2} \rho_{\rm ISM} v_{\rm ISM}^2 \nonumber
\end{eqnarray}
where $\rho_{\rm ISM}$ and $v_{\rm ISM}$ are the mass density and
velocity of the ambient ISM.
The above differential proper-motion analysis yielded $v_{\rm ISM} = 18$
and $109$ km s$^{-1}$ for the shell at 1 and 6 kpc away.
Assuming the ISM number density of 1 cm$^{-3}$ and the average ISM
particle mass of $1.4$ hydrogen atomic mass, $p_{\rm ISM} = 3.9 \times
10^{-12}$ and $1.4 \times 10^{-10}$ dyn cm$^{-2}$ for the shell at 1 and
6 kpc, respectively.
Thus, the internal pressure of the shell is much higher than the ambient
ISM pressure even if the shell is 6 kpc away, and thus, the shell would
be able to retain its spherical structure even with the presence of the
ISM flow local to HD 179821. 

Still, it remains puzzling as to why HD 179821's shell does not show
any apparent signature of ram pressure stripping, which is expected from
the interaction between the shell and the suspected ISM flow.
The strengths of the interaction between the shell and the ISM flow are
dictated principally by the {\sl relative} velocity of the ISM in the
frame of the shell (i.e.\ shock velocity).
We have determined above that the ISM flows only in the tangential
direction with respect to the local standard of rest, (as it is unlikely that an additional line-of-sight radial component of the ISM flow exists due to the symmetry of observed in the five separate CO observations
(\citealt{zd86,likkel87,vtw93,fong06,cas07})). 
In the tangential frame of the stellar wind facing against the ISM flow, the
relative ISM flow velocity amounts to 53 to 70 km s$^{-1}$ and 144 to
245 km s$^{-1}$ respectively at 1 and 6 kpc.  It is unlikely, but if an additional radial component exists, 
the magnitude of the relative ISM flow will be larger; hence, these velocities are the lower limits.

This implies, therefore, that a smaller distance to HD 179821 is
preferred in order for the interactions between the stellar wind and ISM to remain benign, i.e.
to avoid detection by the present and other observations.
This implication is also consistent with the expectation that HD 179821
is located sufficiently closer so that there is enough ISM around the
star with which it could interact.
Given the Galactic latitude of the star ($-5^{\circ}$), HD 179821 is
located roughly 500 pc off the Galactic plane at 6 kpc, which would be
several times the median scale-height of the Milky Way. 
It seems unlikely that the ISM flow at 500 pc above the Galactic plane
could affect the massive circumstellar shell ($M_{\rm sh} \sim 4$
M$_{\odot}$ at 6 kpc) in the way the present investigation has
revealed. 

Hence, this object appears more likely to be a post-AGB object having
some high-mass star characteristics, rather than a high-mass star having
a post-AGB characteristics.
HD 179821 could therefore be an example of a super-AGB star (e.g.\
\citealt{herwig05}), and for that, it probably warrants closer 
investigations into its true nature.
While our analysis did not yield a decisive conclusion for the distance
to this enigmatic star, HD 179821, it uncovered yet another
observational trait that needs to be dealt with in understanding the
true nature of the object.
In any event, the distance is still the key quantity to resolve the
mystery of HD 179821. 
VLBI astrometry may be worthwhile to make a break-through in this
pursuit.

\section{Summary}

Based on the differential proper-motion study of the circumstellar shell 
structure of HD 179821 using two-epoch {\sl HST} images taken 10 yr
apart, we have found that 
\begin{enumerate}
 \item the entire shell has experienced the bulk
       motion at the rate of (2.41$\pm$0.43, 2.97$\pm$0.32) mas yr$^{-1}$,
 \item the shell's bulk motion, combined with proper motion of the
       star, suggests a tangential ISM flow local to HD 179821 
       at the rate of (1.28$\pm$0.95, 7.27$\pm$0.75) mas yr$^{-1}$ at
       most, 
 \item the internal pressure of the shell is orders of
       magnitude higher than the ambient ISM  pressure, and hence, the  
       shell can maintain its symmetric shape while it is being
       blown by the suspected ISM flow,
 \item if the suggested ISM flow local to HD 179821 exists, a smaller
       value for the distance to the object is preferred, implying that
       this object is a post-AGB star with high-mass star
       characteristics. 
\end{enumerate}

\acknowledgements

This research is based on observations with the NASA/ESA {\sl Hubble
Space Telescope}, obtained at the STScI, which is operated by the AURA
under NASA contract NAS 5-26555. 
The authors acknowledge financial support by NASA HST-GO-10837.02-A and
the University of Denver's Professional Research Opportunities for
Faculty grant.
The authors also thank Robert Stencel and Paul Hemenway for interesting
discussions on the topic, especially on the astrometric capability of
{\sl HST}.
Useful comments from the anonymous referee are also appreciated.  
This project's completion is owed to the wisdom, counsel, and mentorship of Dr. Toshiya Ueta.

\end{document}